# A unified stochastic particle method based on the Bhatnagar-Gross-Krook model for polyatomic gases and its combination with DSMC

Fei Fei[1,2,*], Yuan Hu[3,*], Patrick Jenny[2]

1. School of Aerospace Engineering, Huazhong University of science and technology, 430074 Wuhan, China
2. Institute of Fluid Dynamics, ETH Zürich, Sonneggstrasse 3, CH-8092 Zürich, Switzerland
3. State Key Laboratory of High Temperature Gas Dynamics, Institute of Mechanics, Chinese Academy of Sciences, 100190 Beijing, China

\* Corresponding author: ffei@hust.edu.cn; yhu@imech.ac.cn

**Abstract.** Simulating hypersonic flow around a space vehicle is challenging because of the multiscale and nonequilibrium nature inherent in these flows. To effectively deal with such flows, a novel particle-particle hybrid scheme combining the stochastic particle Bhatnagar-Gross-Krook (BGK) method with Direct Simulation Monte Carlo (DSMC) was developed recently, but only for monatomic gases [Fei et. al., J. Comput. Phys. 2021]. Here this work is extended to the particle-particle hybrid method for polyatomic gases. In the near continuum regime, employing the Ellipsoidal–Statistical BGK model proposed by Y. Dauvois, et. al. [Eur. J. Mech. B Fluids, 2021] with discrete levels of vibrational energy, the stochastic particle BGK method is first established following the idea of the unified stochastic particle BGK (USP-BGK) scheme. In the fluid limit, it has been proven to be of second-order temporal and spatial accuracy. Then, the USP-BGK scheme with rotational and vibrational energies is combined with DSMC to construct a hybrid method for polyatomic gases. The present hybrid scheme is validated with numerical tests of homogenous relaxation, 1D shock structure and 2D hypersonic flows past a wedge and a cylinder. Compared to traditional stochastic particle methods, the proposed hybrid method can achieve higher accuracy at a much lower computational cost. Therefore, it is a more efficient tool to study multiscale hypersonic flows.

# 1. Introduction

Hypersonic flows about space vehicles during reentry in near space always exhibit multiscale and nonequilibrium phenomena [1]. For the vibrational thermal nonequilibrium, traditional computational fluid dynamics (CFD) methods can still be employed most of the time, if vibrational modes are modelled by an additional energy equation [2]. However, for translational and rotational thermal nonequilibrium, the continuum equations break down as the Knudsen number becomes relatively large. In this rarefied regime, instead of the continuum equations, the Boltzmann equation describing the details of molecular motion and collisions is required. However, due to the complexity of the collision operator and the high dimensionality of the solution domain, solving the Boltzmann equation numerically is not easy. A popular Boltzmann solver for hypersonic flows is direct simulation Monte Carlo (DSMC) proposed by Bird [3]. Using stochastic particles, the DSMC method can dramatically reduce the difficulty of high dimensionality. However, its efficiency in the continuum regime is quite low due to time step and cell size restrictions dictated by the mean collision time and mean free path. Therefore, simulating multiscale flows with DSMC alone is computationally very expensive.

These time step and cell size restrictions of conventional DSMC (as discussed in [3]) are due to two problems: the stiffness of the Boltzmann collision term [4] and the decoupling of molecular motion and collision in the implementations [5]. For the first problem, certain modifications already exist which improve the collision calculation for near continuum flows [4, 6-7]. From the viewpoint of kinetic models, one also can circumvent the complexity of the Boltzmann collision term using simplified kinetic models, such as the Bhatnagar–Gross–Krook (BGK) [8] and Fokker-Planck (FP) models [9]. These simplified kinetic models can lead to very efficient stochastic particle methods, which employ similar data and code structures as DSMC [10, 11]. The BGK model was first proposed to model monoatomic gas flow and then was extended for applications that include rotational and vibrational energies [12-17]. As the original BGK model assumes a constant Prandtl number equal to one, two kinds of modified

BGK models were introduced to ensure correct transport properties, such as the Ellipsoidal–Statistical BGK model (ES-BGK) [18] and the Shakhov BGK model [19]. Correspondingly, the BGK models for polyatomic gases can be categorized into these two types [20, 21]. Superior to the Shakhov model, according to our knowledge, only the ES-BGK model was proven to satisfy the H-theorem in kinetic theory [21]. Recently, an ES-BGK model with discrete levels of vibrational energy was proposed by Y. Dauvois, et. al. [22], which satisfies the H-theorem and corrects Prandtl number and relaxation times of rotational and vibrational energies. Most of the BGK models were constructed by assuming a continuous distribution of the vibrational energy. However, with discrete vibrational energies, Dauvois' ES-BGK model can better approach the physics of hypersonic flows. Therefore, it is selected in the present work. In addition, it is worthwhile to mention that the FP model has also been extended for polyatomic gases by Gorji and Jenny [23] and by Mathiaud and Mieussens [24, 25].

If the stochastic particle methods based on the BGK and FP models were implemented in a similar way as conventional DSMC, i.e., by splitting molecular motion and collisions, their accuracy would reduce to the first-order in the fluid limit [5, 26]. Therefore, to avoid strong numerical dissipation, time step and cell sizes in traditional stochastic particle methods are severely restricted. Recently, a modified stochastic particle method named the unified stochastic particle BGK method (USP-BGK) was developed [26] to overcome this defect. In the USP-BGK method, the collision term of the BGK model was decomposed into continuum and rarefied parts. By resolving the continuum part with a high order scheme, the USP-BGK method can achieve second-order accuracy. In this paper, following the strategy of this micro-macro decomposition of the collision term, the previous USP-BGK method is extended to deal with polyatomic gas flows.

As well known, both BGK and FP models only are applicable when the Knudsen number is not too large [27]. For multiscale flows with a wide range of Knudsen numbers, a simple choice is to combine a stochastic particle method of the BGK or FP type with DSMC. Different hybrid methods already exist, in which DSMC is coupled with BGK [28, 29] or FP [30-33] solvers. Compared to the CFD-DSMC hybrid schemes [34–36], particle-particle hybrid methods are free of the strict requirements on positioning the hybrid interface and no instabilities occur due to the coupling of deterministic and stochastic algorithms [37]. As a natural extension of the proposed USP-BGK method, such a hybrid scheme for polyatomic gas flow also is devised in the current paper. It is shown that the new hybrid algorithm achieves higher efficiency and accuracy than traditional BGK-DSMC hybrid methods.

Our paper is organized as follows. In section 2, the ES-BGK model for polyatomic gases in ref. [22] is reviewed. Based on this ES-BGK model, a traditional stochastic particle method is introduced in section 3. After that, the unified stochastic particle BGK method for polyatomic gases and a corresponding hybrid schene in combination with DSMC are proposed and analyzed in sections 4 and 5, respectively. At last, section 6 provides numerical validations for several typical multiscale gas flows.

## 2. Review of the ES-BGK model for polyatomic gases

The Bhatnagar-Gross-Krook (BGK) model simplifies the Boltzmann collision term using a relaxation process. Recently, Y. Dauvois, et. al. [22] proposed an ES-BGK model with discrete levels of vibrational energy for polyatomic gases. The kinetic equation of this ES-BGK model is given by

$$\frac{\partial \mathcal{F}}{\partial t} + V_i \frac{\partial \mathcal{F}}{\partial x_i} = Q(\mathcal{F}) = \frac{1}{\tau_{BGK}}(\mathcal{F}_G - \mathcal{F}), \qquad (2.1)$$

where $Q(\mathcal{F})$ denotes the BGK collision term and $\mathcal{F}(\mathbf{V}, I_r, I_v; \mathbf{x}, t) = \rho(\mathbf{x}, t) f(\mathbf{V}, I_r, I_v; \mathbf{x}, t)$ is the mass density function depending on time $t$ and position $\mathbf{x}$. Further, $\rho$ is the gas density and $f$ the joint probability density function (PDF) of the molecular velocity $\mathbf{V}$, rotational energy $I_r$ and vibrational energy $I_v$. $\tau_{BGK}$ denotes the relaxation time of the ES-BGK model. Since a single relaxation time is not enough to account for the correct Prandtl number as well as the relaxation rate of different internal energies, in the ES-BGK model proposed in ref. [22] a Gaussian distribution $\mathcal{F}_G$ was introduced to replace the Maxwellian distribution, i.e.,

$$\mathcal{F}_G(\mathbf{V}, I_r, I_v) = \rho g_{tr}(\mathbf{V}) g_{rot}(I_r) g_{vib}(I_v), \tag{2.2}$$

with

$$g_{tr}(\mathbf{V}) = \frac{1}{\sqrt{\det(2\pi\Pi)}} \exp\left(-\frac{1}{2}(\mathbf{V}-\mathbf{U})^T \Pi^{-1}(\mathbf{V}-\mathbf{U})\right), \tag{2.3a}$$

$$g_{rot}(I_r) = \frac{\Lambda(\delta_{rot})}{(RT_{rot}^{rel})^{\delta_{rot}/2}} I_r^{\frac{\delta_{rot}-2}{2}} \exp\left(-\frac{I_r}{RT_{rot}^{rel}}\right) \quad \text{and} \tag{2.3b}$$

$$g_{vib}(I_v) = \left[1-\exp\left(-\frac{T_0}{T_{vib}^{rel}}\right)\right] \exp\left(-\frac{I_v}{RT_{vib}^{rel}}\right), \tag{2.3c}$$

where $g_{tr}(\mathbf{V})$, $g_{rot}(I_r)$ and $g_{vib}(I_v)$ are distributions associated with the translational, rotational and vibrational energies of the gaseous molecules, $\Lambda(\delta) = 1/\Gamma(\delta/2)$, and $\Gamma$ is the usual gamma function. For clarity, functions $E_i(\mathcal{F}) = e_i(T)$ that map translational, rotational and vibrational temperature to the corresponding energies are defined in ref. [22], i.e.,

$$E_{tr}(\mathcal{F}) = e_{tr}(T_{tr}) = \frac{3}{2} RT_{tr}, \tag{2.4a}$$

$$E_{rot}(\mathcal{F}) = e_{rot}(T_{rot}) = \frac{\delta_{rot}}{2} RT_{rot} \quad \text{and} \tag{2.4b}$$

$$E_{vib}(\mathcal{F}) = e_{vib}(T_{vib}) = \frac{RT_0}{\exp(T_0/T_{vib})-1}, \tag{2.4c}$$

where $\delta_{rot}$ is the number of degrees of freedom of rotation, and $T_0$ is the characteristic vibrational temperature. Moreover, the intermediate translational-rotational temperature $T_{tr,rot}$ and equilibrium temperature $T_{eq}$ are defined as

$$e_{tr,rot}(T_{tr,rot}) = E_{tr}(\mathcal{F}) + E_{rot}(\mathcal{F}) \quad \text{and} \tag{2.5a}$$

$$e(T_{eq}) = E_{tr}(\mathcal{F}) + E_{rot}(\mathcal{F}) + E_{vib}(\mathcal{F}). \tag{2.5b}$$

According to Eqs. (2.4) and (2.5), related temperatures can be obtained from the inverse functions

$$T_{tr} = e_{tr}^{-1}(E_{tr}), \quad T_{rot} = e_{rot}^{-1}(E_{rot}) \quad \text{and} \quad T_{vib} = e_{vib}^{-1}(E_{vib}), \tag{2.6}$$

and

$$T_{tr,rot} = e_{tr,rot}^{-1}(E_{tr} + E_{rot}) \quad \text{and} \quad T_{eq} = e^{-1}(E_{tr} + E_{rot} + E_{vib}). \tag{2.7}$$

The translational, rotational and vibrational energies of the gas can be obtained from the ensemble average as

$$\rho E_{tr}(\mathcal{F}) = \left\langle \frac{1}{2}(V-U)^2 \mathcal{F} \right\rangle_{V,I_r,I_v}, \quad \rho E_{rot}(\mathcal{F}) = \left\langle I_r \mathcal{F} \right\rangle_{V,I_r,I_v} \quad \text{and} \quad \rho E_{vib}(\mathcal{F}) = \left\langle I_v \mathcal{F} \right\rangle_{V,I_r,I_v}, \tag{2.8}$$

where $\mathbf{U}$ is the mean velocity. For the discrete vibrational modes $I_v = iRT_0$ and $\langle \phi \rangle_{V,I_r,i} = \sum_{i=0}^{+\infty} \int_{\mathbb{R}^3} \int_{\mathbb{R}} \phi(\mathbf{V}, I_r, i) dI_r d\mathbf{V}$ denotes the integral of any function $\phi$.

In the Gaussian distribution (2.2) the covariance matrix $\Pi$, the temperatures $T_{rot}^{rel}$ and $T_{vib}^{rel}$ are employed to fit different the relaxation times given by [22], i.e.,

$$\Pi = \eta RT_{eq}\mathbf{I} + (1-\eta)\left[\theta RT_{tr,rot}\mathbf{I} + (1-\theta)(\nu\Theta + (1-\nu)RT_{tr}\mathbf{I})\right], \tag{2.9}$$

$$e_{rot}^{rel} = \eta e_{rot}(T_{eq}) + (1-\eta)\left[\theta e_{rot}(T_{tr,rot}) + (1-\theta)E_{rot}(\mathcal{F})\right], \tag{2.10}$$

$$e_{vib}^{rel} = \eta e_{vib}(T_{eq}) + (1-\eta)E_{vib}(\mathcal{F}), \tag{2.11}$$

where $\eta$, $\theta$ and $\nu$ are relaxation coefficients. $\Theta$ is related to a pressure tensor, i.e.,

$$\rho\Theta = \left\langle (\mathbf{V}-\mathbf{U}) \otimes (\mathbf{V}-\mathbf{U})\mathcal{F} \right\rangle_{V,I_r,I_v}, \tag{2.12}$$

and $T_{rot}^{rel}$ and $T_{vib}^{rel}$ are calculated from $e_{rot}^{rel}$ and $e_{vib}^{rel}$ as

$$T_{rot}^{rel} = e_{rot}^{-1}(e_{rot}^{rel}) \quad \text{and} \quad T_{vib}^{rel} = e_{vib}^{-1}(e_{vib}^{rel}). \tag{2.13}$$

The relaxation coefficient $\theta$ is associated with the transfer between translational and rotational energies and $\eta$ refers to the transformation between translational-rotational and vibrational energies. As shown in ref. [22] we have

$$\eta = \frac{1}{Z_{vib}} \quad \text{and} \quad \theta = \frac{1/Z_{rot} - 1/Z_{vib}}{1 - 1/Z_{vib}}, \tag{2.14}$$

where $Z_{rot}$ and $Z_{vib}$ are the collision numbers describing the relaxation rate of the rotational and vibrational energies to equilibrium. Assuming $\tau_{rot}$ and $\tau_{vib}$ as their relaxation times, one obtains

$$\tau_{rot} = \tau_{BGK} Z_{rot} \quad \text{and} \quad \tau_{vib} = \tau_{BGK} Z_{vib}. \tag{2.15}$$

The relaxation coefficient $\nu$ is used to fit the correct Pr number given by

$$\Pr = \frac{1}{1-(1-\eta)(1-\theta)\nu}. \tag{2.16}$$

Finally, the relaxation time $\tau_{BGK}$ of the ES-BGK model is

$$\tau_{BGK} = \frac{\mu}{p}\left[1-(1-\eta)(1-\theta)\nu\right], \tag{2.17}$$

where $\mu$ and $p$ are gas viscosity and pressure, respectively.

Appling the aforementioned ES-BGK model for polyatomic gases, a unified stochastic particle method will be developed to deal with multiscale hypersonic flows.

## 3. Traditional stochastic particle method based on the ES-BGK model

To solve the kinetic equation (2.1) numerically, a stochastic particle method based on the ES-BGK model was first proposed by Gallis and Torczynski [10] for monoatomic gases and then extended to include internal energies [38-40]. These traditional stochastic particle methods employ a simple time splitting scheme. Therefore, the governing equation (2.1) is divided into a transport step and a relaxation step, i.e.,

**transport step:** $\quad \dfrac{\partial \mathcal{F}}{\partial t} + V_i \dfrac{\partial \mathcal{F}}{\partial x_i} = 0 \quad$ and $\tag{3.1a}$

**relaxation step:** $\quad \dfrac{\partial \mathcal{F}}{\partial t} = Q(\mathcal{F}). \tag{3.1b}$

The transport step is updated by moving the computational particles with their molecular velocities. Thus the solution of Eq. (3.1a) can be obtained as

$$\mathcal{F}^*(\mathbf{V}, I_r, I_v; \mathbf{x}, \Delta t) = \mathcal{F}(\mathbf{V}, I_r, I_v; \mathbf{x}-\mathbf{V}\Delta t, t^n), \tag{3.2}$$

where $\mathcal{F}^*$ is the intermediate distribution function after particle movement. Then, setting $\mathcal{F}^*$ as the initial condition, the distribution function after the relaxation step is calculated as

$$\mathcal{F}(\mathbf{V}, I_r, I_v; \mathbf{x}, t^{n+1}) = \mathcal{F}^*(\mathbf{V}, I_r, I_v; \mathbf{x}, \Delta t) e^{-\Delta t/\tau_{BGK}} \\ + \left(1-e^{-\Delta t/\tau_{BGK}}\right) \int_{t^n}^{t^n+\Delta t} \frac{e^{t/\tau_{BGK}}}{\tau_{BGK}\left(e^{\Delta t/\tau_{BGK}}-1\right)} \mathcal{F}_G(\mathbf{V}, I_r, I_v; \mathbf{x}, t) dt. \tag{3.3}$$

Generally, the Gaussian distribution $\mathcal{F}_G(\mathbf{V}, I_r, I_v; \mathbf{x}, t)$, which depends on $\mathbf{\Pi}(t)$, $E_{tr}(t)$, $E_{rot}(t)$ and $E_{vib}(t)$, is not constant during the relaxation process. The solution of $\mathbf{\Pi}(t)$, $E_{tr}(t)$, $E_{rot}(t)$ and $E_{vib}(t)$ at a certain time $t$ can be calculated using the moment equation (2.1) (see Appendix A).

From the viewpoint of computational particles, Eq. (3.3) can be implemented as follows: first, a random number $rand$ between 0 and 1 is sampled and compared with $e^{-\Delta t / \tau_{BGK}}$. If $rand < e^{-\Delta t / \tau_{BGK}}$, the velocities, rotational and vibrational energies of the selected particle remain unchanged; otherwise, according to the third term on the right-hand side of Eq. (3.3), they need to be resampled from the distribution $\mathcal{F}_G(\mathbf{V}, I_r, I_v; \mathbf{x}, t)$. A certain time $t$ in the range $[0, \Delta t]$ is calculated as

$$t = \tau_{BGK} \ln\left[ Rf \left( e^{\Delta t / \tau_{BGK}} - 1 \right) + 1 \right] \tag{3.4}$$

using Monte Carlo, where $Rf$ is a random number between 0 and 1. After the time $t$ is known, $\mathcal{F}_G(\mathbf{V}, I_r, I_v; \mathbf{x}, t)$ can be determined from the values of $\mathbf{\Pi}(t)$, $E_{tr}(t)$, $E_{rot}(t)$ and $E_{vib}(t)$ (see Appendix A). A Metropolis-Hastings (MH) method is employed to sample the velocities, rotational and vibrational energies from the distribution $\mathcal{F}_G(\mathbf{V}, I_r, I_v; \mathbf{x}, t)$ (details of the MH algorithm are given in Appendix B).

The implementation of the traditional stochastic particle BGK method is outlined in Table 1.

**Table 1.** Outline of the implementation of the traditional stochastic particle BGK method

| | |
|---|---|
| 1. **Initialization** | Introduce initial computational particles in the computational domain. Their velocities, rotational and vibrational energies are sampled from the initial distribution. |
| 2. **Transport** | Move the computational particles with their velocities and apply boundary conditions to obtain $\mathcal{F}^*(\mathbf{V}, I_r, I_v; \mathbf{x}, \Delta t)$. |
| 3. **Relaxation** | $N\left(1 - e^{-\Delta t / \tau_{BGK}}\right)$ computational particles are randomly selected to assign |

new values sampled from the PDF $\mathcal{F}_G(\mathbf{V}, I_r, I_v; \mathbf{x}, t)$, where the time t is sampled from Eq. (3.4) and related mean variables are calculated as explained in Appendix A; the values of the remaining particles remain unchanged.

4. **Sampling**     Sample the macroscopic quantities from the computational particles.

## 4. The USP-BGK method for polyatomic gases

By introducing characteristic variables, such as $x_*$, $\rho_*$ and $V_* = \sqrt{8RT_*/\pi}$, the BGK collision term can be rewritten in dimensionless form, i.e.,

$$Q(\mathcal{F}) = \frac{1}{\varepsilon}(\mathcal{F}_G - \mathcal{F}), \tag{4.1}$$

where $\varepsilon = \tau_{BGK} V_* / x_*$ denotes the Knudsen number. It is noted that the BGK equation shows strong stiffness in the continuum regime. Since the traditional stochastic particle methods (SP-BGK) employ the time splitting scheme, this stiffness of the BGK collision term will reduce the order of accuracy in the fluid limit [5]. Therefore, traditional SP-BGK methods turn to first order in the fluid limit and suffer from low accuracy and efficacy. To overcome this defect, a unified stochastic particle BGK method (USP-BGK) for monatomic gases was developed recently [26]. In this work, the USP-BGK method is further extended for polyatomic gases.

**Remark 4.1:** We refer to the fluid limit if $\Delta t / \varepsilon \gg 1$, and to the kinetic limit if $\Delta t / \varepsilon \ll 1$, where $\Delta t$ denotes the time step size. However, the definitions of the continuum and rarefied regimes are independent of the time step size, i.e., one refers to the continuum regime if $\varepsilon \ll 1$ and to the rarefied regime if $\varepsilon \gg 1$.

4.1 Governing equations

Similar to the USP-BGK method for monatomic gases [26] we decompose the ES-BGK collision term for polyatomic gases into continuum and rarefied parts, i.e.,

$$Q(\mathcal{F}) = Q_C(\mathcal{F}) + Q_R(\mathcal{F}), \tag{4.2}$$

The continuum part $Q_C(\mathcal{F})$ is closed by the Grad expansion of the molecular PDF with low order moments, i.e.,

$$Q_C(\mathcal{F}) = \frac{1}{\varepsilon}\left(\mathcal{F}_M - \mathcal{F}_{|Grad}\right). \tag{4.3}$$

$Q_R(\mathcal{F})$ denotes the remaining part of the BGK collision term, i.e., the rarefied part given by

$$Q_R(\mathcal{F}) = Q(\mathcal{F}) - Q_C(\mathcal{F}) = \frac{1}{\varepsilon}(\mathcal{F}_G - \mathcal{F}) - Q_C(\mathcal{F}). \tag{4.4}$$

In Eq. (4.3), the Maxwellian distribution $\mathcal{F}_M$ is given by

$$\mathcal{F}_M(\mathbf{V}, I_r, I_v) = \rho M_{tr}(\mathbf{V}) M_{rot}(I_r) M_{vib}(I_v) \tag{4.5}$$

with

$$M_{tr}(\mathbf{V}) = \frac{1}{(2\pi RT_{tr})^{3/2}} \exp\left(-\frac{(\mathbf{V}-\mathbf{U})^2}{2RT_{tr}}\right), \tag{4.6a}$$

$$M_{rot}(I_r) = \frac{\Lambda(\delta_{rot})}{(RT_{rot})^{\delta_{rot}/2}} I_r^{\frac{\delta_{rot}-2}{2}} \exp\left(-\frac{I_r}{RT_{rot}}\right) \text{ and} \tag{4.6b}$$

$$M_{vib}(I_v) = \left[1 - \exp\left(-\frac{T_0}{T_{vib}}\right)\right] \exp\left(-\frac{I_v}{RT_{vib}}\right). \tag{4.6c}$$

The Grad distribution $\mathcal{F}_{|Grad}$ is constructed from the low order moments of the Grad distribution, such as the density, mean velocity, shear stress and heat flux, together with the translational, rotational and vibrational energies. The expansion coefficients of the Grad distribution are calculated to ensure that the low order moments of the continuum part are equal to those of the original collision term, i.e.,

$$\langle \phi Q_C(\mathcal{F}) \rangle_{V, I_r, I_v} = \langle \phi Q(\mathcal{F}) \rangle_{V, I_r, I_v}, \tag{4.7}$$

where

$$\phi = \left(1, \mathbf{V}, \frac{1}{2}(V-U)^2, I_r, I_v, (\mathbf{V}\cdot\mathbf{U})\otimes(\mathbf{V}\cdot\mathbf{U}) - RT_{tr}\mathbf{I}, \frac{1}{2}(\mathbf{V}\cdot\mathbf{U})(V-U)^2, \frac{1}{2}(\mathbf{V}\cdot\mathbf{U})I_r, \frac{1}{2}(\mathbf{V}\cdot\mathbf{U})I_v\right)^T,$$

and the heat flux related to the translational ($\mathbf{q}_{tr}$), rotational ($\mathbf{q}_{rot}$) and vibrational ($\mathbf{q}_{vib}$) energies are also separated as the last three-term in the above bracket. Therefore $Q_C$ is obtained as

$$Q_C = -\frac{1}{\varepsilon}\mathcal{F}_M \left\{ \begin{array}{l} \dfrac{1}{2\rho RT_{tr}\,\mathrm{Pr}}\left[\dfrac{\mathbf{C}\otimes\mathbf{C}}{RT_{tr}} - \dfrac{1}{3}\dfrac{C^2}{RT_{tr}}\mathbf{I}\right]:\boldsymbol{\sigma} \\ + \dfrac{2}{5\rho RT_{tr}}\left(\dfrac{C^2}{2RT_{tr}} - \dfrac{5}{2}\right)\dfrac{\mathbf{C}}{RT_{tr}}\cdot\mathbf{q}_{tr} + \dfrac{2}{\delta_{rot}\rho RT_{rot}}\left(\dfrac{I_r}{RT_{rot}} - \dfrac{\delta_{rot}}{2}\right)\dfrac{\mathbf{C}}{RT_{tr}}\cdot\mathbf{q}_{rot} \\ + \dfrac{1}{\rho T_{vib}\dfrac{\partial e_{vib}}{\partial T_{vib}}}\left(I_v\dfrac{T_0}{T_{vib}} - \dfrac{\delta_v(T_{vib})}{2}\right)\dfrac{\mathbf{C}}{RT_{tr}}\cdot\mathbf{q}_{vib} \\ -\dfrac{1}{T_{tr}}\left(\dfrac{C^2}{2RT_{tr}} - \dfrac{3}{2}\right)\left[\eta(T_{eq}-T_{tr}) + (1-\eta)\left(\dfrac{3\theta}{3+\delta_{rot}}E_{rot} - \dfrac{\theta\delta_{rot}}{3+\delta_{rot}}E_{tr}\right)\right]/\left(\dfrac{3}{2}R\right) \\ -\dfrac{1}{T_{rot}}\left(\dfrac{I_r}{RT_{rot}} - \dfrac{\delta_{rot}}{2}\right)\left[\eta(T_{eq}-T_{rot}) + (1-\eta)\left(-\dfrac{3\theta}{3+\delta_{rot}}E_{rot} + \dfrac{\theta\delta_{rot}}{3+\delta_{rot}}E_{tr}\right)\right]/\left(\dfrac{\delta_{rot}}{2}R\right) \\ -\dfrac{\eta}{T_{vib}}\left(I_v\dfrac{T_0}{T_{vib}} - \dfrac{\delta_{vib}(T_{vib})}{2}\right)\left[e_{vib}(T_{eq}) - E_{vib}\right]/\left(\dfrac{\partial e_{vib}}{\partial T_{vib}}\right) \end{array}\right\}$$

, (4.8)

where $\dfrac{\partial e_{vib}(T_{vib})}{\partial T_{vib}} = R\dfrac{T_0^2}{T_{vib}^2}\left\{\exp\left(\dfrac{T_0}{T_{vib}}\right)\bigg/\left[\exp\left(\dfrac{T_0}{T_{vib}}\right)-1\right]^2\right\}$ , $\delta_{vib}(T_{vib}) = \dfrac{2T_0/T_{vib}}{\exp(T_0/T_{vib})-1}$ denotes the number of vibrational degrees of freedom and $\boldsymbol{\sigma}$ is the shear stress.

Based on this decomposition of the BGK collision term, the governing equations for the unified stochastic particle BGK method is constructed as follows:

**transport step:** $\dfrac{\partial \mathcal{F}}{\partial t} + V_i \dfrac{\partial \mathcal{F}}{\partial x_i} = Q_C(\mathcal{F})$ and (4.9a)

**relaxation step:** $\dfrac{\partial \mathcal{F}}{\partial t} = Q_R(\mathcal{F})$. (4.9b)

In the USP-BGK method, the transport and relaxation steps are implemented in a sequence similar to that of the traditional SP-BGK method.

For the transport step, the trapezoidal rule is used to approach the integral of the relaxation term. Therefore the numerical solution of the transport step along characteristics is given by

$$\mathcal{F}^*(\mathbf{V}, I_r, I_v; \mathbf{x}, \Delta t) = \dfrac{\Delta t}{2}\left[Q_C^*(\mathbf{V}, I_r, I_v; \mathbf{x}, \Delta t) + Q_C(\mathbf{V}, I_r, I_v; \mathbf{x}-\mathbf{V}\Delta t, t^n)\right]$$
$$+ \mathcal{F}(\mathbf{V}, I_r, I_v; \mathbf{x}-\mathbf{V}\Delta t, t^n)$$
(4.10)

For the relaxation step, taking the intermediate PDF $\mathcal{F}^*(\mathbf{V}, I_r, I_v; \mathbf{x}, \Delta t)$ as the initial condition and calculating the integral solution of Eq. (4.9b), the final PDF solution is obtained as

$$\mathcal{F}(\mathbf{V},I_r,I_v;\mathbf{x},t^{n+1}) = \mathcal{F}^*(\mathbf{V},I_r,I_v;\mathbf{x},\Delta t)e^{-\Delta t/\varepsilon}$$
$$+\left(1-e^{-\Delta t/\varepsilon}\right)\int_{t^n}^{t^n+\Delta t}\frac{e^{t/\varepsilon}}{\varepsilon\left(e^{\Delta t/\varepsilon}-1\right)}\mathcal{F}_G(\mathbf{V},I_r,I_v;\mathbf{x},t)dt - \varepsilon\left(1-e^{-\Delta t/\varepsilon}\right)\cdot Q_C^*(\mathbf{V},I_r,I_v;\mathbf{x},\Delta t). \quad (4.11)$$

## 4.2 The simplified algorithm of the unified stochastic particle method

Using stochastic particles, the governing equations (4.10) and (4.11) can be numerically solved following the simplified algorithm of the USP-BGK method [29, 41]. By introducing the two auxiliary PDFs

$$\tilde{\mathcal{F}}^* = \mathcal{F}^* - \frac{\Delta t}{2}Q_C^* \quad \text{and} \quad (4.12a)$$

$$\hat{\mathcal{F}} = \mathcal{F} + \frac{\Delta t}{2}Q_C \quad (4.12b)$$

the solution of transport step (4.10) can be arranged as

$$\tilde{\mathcal{F}}^*(\mathbf{V},I_r,I_v;\mathbf{x},\Delta t) = \hat{\mathcal{F}}(\mathbf{V},I_r,I_v;\mathbf{x}-\mathbf{V}\Delta t,t^n). \quad (4.13)$$

Similarly, using $\tilde{\mathcal{F}}^*$ and $\hat{\mathcal{F}}$, the solution of the relaxation step (4.11) can be rewritten as

$$\hat{\mathcal{F}}(\mathbf{V},I_r,I_v;\mathbf{x},t^{n+1}) = \tilde{\mathcal{F}}^*(\mathbf{V},I_r,I_v;\mathbf{x},\Delta t)e^{-\Delta t/\varepsilon} + \left(1-e^{-\Delta t/\varepsilon}\right)\mathcal{F}_U^*(\mathbf{V},I_r,I_v;\mathbf{x},\Delta t), \quad (4.14)$$

where a new target distribution for the simplified unified stochastic particle algorithm is defined, i.e.,

$$\mathcal{F}_U^*(\mathbf{V},I_r,I_v;\mathbf{x},\Delta t) = \mathcal{F}_G^*(\mathbf{V},I_r,I_v;\mathbf{x},\Delta t) + \left[\frac{\Delta t}{2}\frac{\left(1+e^{-\Delta t/\varepsilon}\right)}{\left(1-e^{-\Delta t/\varepsilon}\right)} - \varepsilon\right]Q_C^*(\mathbf{V},I_r,I_v;\mathbf{x},\Delta t). \quad (4.15)$$

In the above derivation, the equality of the low order moments of the continuum part and the original collision term was employed. Therefore we have $\mathcal{F}_G(\mathbf{V},I_r,I_v;\mathbf{x},t) = \mathcal{F}_G^*(\mathbf{V},I_r,I_v;\mathbf{x},\Delta t)$ and $Q_C^*(\mathbf{V},I_r,I_v;\mathbf{x},\Delta t) = Q_C(\mathbf{V},I_r,I_v;\mathbf{x},t^{n+1})$.

By comparing Eqs. (4.13) and (4.14) with Eqs. (3.2) and (3.3), the evolution of the original PDF $\mathcal{F}$ in the traditional SP-BGK method can be replaced by updating the auxiliary PDF $\hat{\mathcal{F}}$ in the USP-BGK method. Therefore, the simplified algorithm of the USP-BGK method can be implemented in the same way as the traditional SP-BGK method.

First, the computational particles are initialized and sampled from the initial distribution $\hat{\mathcal{F}}(\mathbf{V}, I_r, I_v; \mathbf{x}, 0)$. Their values are $\{\mathbf{X}^{(\alpha)}, \mathbf{M}^{(\alpha)}, I_r^{(\alpha)}, I_v^{(\alpha)}\}_{\alpha \in N}$ ($\alpha$ is the particle index), where $\mathbf{X}^{(\alpha)}$, $\mathbf{M}^{(\alpha)}$, $I_r^{(\alpha)}$ and $I_v^{(\alpha)}$ denote position, velocity, rotational and vibrational energies of the particle $\alpha$. Since the initial PDF is mostly assumed Maxwellian, $Q_C = 0$ and $\hat{\mathcal{F}}(\mathbf{V}, I_r, I_v; \mathbf{x}, 0) = \mathcal{F}_M(\mathbf{V}, I_r, I_v; \mathbf{x}, 0)$. Therefore the initialization for the USP-BGK method is as same as for DSMC. Otherwise, the distribution $\hat{\mathcal{F}}(\mathbf{V}, I_r, I_v; \mathbf{x}, 0)$ should be exactly sampled using the Metropolis-Hastings algorithm.

Second, for the transport step the velocity, rotational and vibrational energies of the computational particles are kept constant. Their locations, however, are updated along the characteristics, i.e., $\mathbf{X}^{(\alpha)}(t^{n+1}) = \mathbf{X}^{(\alpha)}(t^n) + \mathbf{M}^{(\alpha)}(t^n)\Delta t$. As a result, the intermediate PDF after the transport step is obtained as shown in Eq. (4.13).

Third, for the relaxation step the computational particles are randomly separated into two sets like in the traditional SP-BGK method. The first set has $e^{-\Delta t/\varepsilon} N$ and the second set $(1 - e^{-\Delta t/\varepsilon})N$ particles. Particle velocity, rotational and vibrational energies of the first set remain unchanged, while those of the second set are updated according to the second term on the right-hand side of Eq. (4.14), i.e., sampled from $\mathcal{F}_U^*(\mathbf{V}, I_r, I_v; \mathbf{x}, \Delta t)$ with the Metropolis-Hastings algorithm. In addition, the macro variables used in the construction of $\mathcal{F}_U^*(\mathbf{V}, I_r, I_v; \mathbf{x}, \Delta t)$ are calculated based on the PDF $\tilde{\mathcal{F}}^*(\mathbf{V}, I_r, I_v; \mathbf{x}, \Delta t)$ after the transport step (see Appendix C).

The implementation of the simplified unified stochastic particle method with internal energies is outlined in Table 2.

Table 2. Outline of the simplified algorithm of the USP-BGK method for polyatomic gases

| | |
|---|---|
| 1. **Initialization** | Introduce initial computational particles in the computational domain. Their velocities, rotational and vibrational energies are sampled from the initial auxiliary PDF $\hat{\mathcal{F}}(\mathbf{V}, I_r, I_v; \mathbf{x}, 0)$. |
| 2. **Transport** | Move the computational particles with their velocities and apply |

boundary conditions to obtain $\tilde{\mathcal{F}}^*(\mathbf{V}, I_r, I_v; \mathbf{x}, \Delta t)$.

3. **Relaxation**      $N(1-e^{-\Delta t/\varepsilon})$ computational particles are randomly selected and new values sampled from the PDF $\mathcal{F}_U^*$ are assigned; the values of the remaining particles remain unchanged. $\mathcal{F}_U^*$ is calculated based on Eq. (4.15). After the relaxation step, the PDF of computational particles equals to $\hat{\mathcal{F}}(\mathbf{V}, I_r, I_v; \mathbf{x}, \Delta t)$.

4. **Sampling**      Sample the macroscopic quantities from the auxiliary PDFs (see Appendix C).

4.3 Anlalysis of the USP-BGK method

Similar to the previous version of the USP-BGK method, the accuracy of the current scheme with internal energies also is second order in the fluid regime. When $\Delta t / \varepsilon \gg 1$, the solution of the USP-BGK method (Eq. (4.11)) approaches

$$\mathcal{F}(\mathbf{V}, I_r, I_v; \mathbf{x}, t^{n+1}) = \mathcal{F}_G^*(\mathbf{V}, I_r, I_v; \mathbf{x}, \Delta t) - \varepsilon Q_C^*(\mathbf{V}, I_r, I_v; \mathbf{x}, \Delta t) \ . \tag{4.16}$$

Using Chapann-Enskog analysis one obtains

$$Q_C^*(\mathbf{V}, I_r, I_v; \mathbf{x}, \Delta t) = \frac{\partial_0 M^*}{\partial t} + \mathbf{V} \cdot \nabla M^* + \mathrm{O}(\varepsilon), \tag{4.17}$$

where $M$ is the Maxwellian distribution defined as

$$M(\mathbf{V}, I_r, I_v) = \rho \frac{1}{(2\pi R T_{eq})^{3/2}} \exp\left(-\frac{(\mathbf{V}-\mathbf{U})^2}{2 R T_{eq}}\right) \\ \cdot \frac{\Lambda(\delta_{rot})}{(RT_{eq})^{\delta_{rot}/2}} I_r^{\frac{\delta_{rot}-2}{2}} \exp\left(-\frac{I_r}{RT_{eq}}\right) \cdot \left[1-\exp\left(-\frac{T_0}{T_{eq}}\right)\right] \exp\left(-\frac{I_v}{RT_{eq}}\right) . \tag{4.18}$$

Thus Eq. (4.16) reads

$$\mathcal{F}(\mathbf{V}, I_r, I_v; \mathbf{x}, t^{n+1}) = \mathcal{F}_G^* - \varepsilon\left(\frac{\partial_0 M^*}{\partial t} + \mathbf{V} \cdot \nabla M^*\right) + \mathrm{O}(\varepsilon^2) \ . \tag{4.19}$$

According to Proposition 6.1 of ref. [22], Eq. (4.19) leads to

$$\mathcal{F}(\mathbf{V}, I_r, I_v; \mathbf{x}, t^{n+1}) = \mathcal{F}^*(\mathbf{V}, I_r, I_v; \mathbf{x}, \Delta t) + \mathrm{O}(\varepsilon^2) \ . \tag{4.20}$$

Substituting Eq. (4.10) into Eq. (4.20), using $Q_C^*(\mathbf{V}, I_r, I_v; \mathbf{x}, \Delta t) = Q_C(\mathbf{V}, I_r, I_v; \mathbf{x}, t^{n+1})$ and

$Q^* = Q_c^* + O(\varepsilon)$, one obtains

$$\mathcal{F}(\mathbf{V}, I_r, I_v; \mathbf{x}, t^{n+1}) = \frac{\Delta t}{2} \left[ Q(\mathbf{V}, I_r, I_v; \mathbf{x}, t^{n+1}) + Q(\mathbf{V}, I_r, I_v; \mathbf{x} - \mathbf{V}\Delta t, t^n) \right] \\ + \mathcal{F}(\mathbf{V}, I_r, I_v; \mathbf{x} - \mathbf{V}\Delta t, t^n) + O(\Delta t \varepsilon) \quad (4.21)$$

Expanding the distribution functions around $\mathbf{x}' = \mathbf{x} - \mathbf{V}\Delta t/2$ and $t' = \Delta t/2$, and after reorganizing, Eq. (4.21) yields

$$\frac{\partial \mathcal{F}}{\partial t} + V_i \frac{\partial \mathcal{F}}{\partial x_i} = Q(\mathcal{F}) + O(\Delta t^2) + O(\varepsilon). \quad (4.22)$$

Opposed to the traditional SP-BGK method, we note that the USP-BGK method for polyatomic gases is of second-order accuracy in the fluid limit.

4.4 Space interpolation and boundary conditions

As shown in Eq. (4.22), the USP-BGK method is of second-order accuracy in time. To also achieve second-order accuracy in space, spatial interpolation of second-order should also be employed. For example, considering two-dimensional cases using Taylor expansion, any macro quantity $\Phi$, such as $\rho$, $\mathbf{U}$ or $E_{tr}$ at the particle location $(x, y)$, can be reconstructed about the cell center $(x_c, y_c)$, i.e.,

$$\Phi(x, y) = \Phi_c + \frac{\partial \Phi_c}{\partial x}(x - x_c) + \frac{\partial \Phi_c}{\partial y}(y - y_c) \\ + \frac{\partial^2 \Phi_c}{\partial x^2}\frac{(x - x_c)^2}{2} + \frac{\partial^2 \Phi_c}{\partial x \partial y}(x - x_c)(y - y_c) + \frac{\partial^2 \Phi_c}{\partial y^2}\frac{(y - y_c)^2}{2} + O(\Delta h^3) \quad (4.23)$$

where $\Phi_c$ and $\partial^{m+l}\Phi_c/\partial x^m \partial y^l$ are unknown coefficients, which respectively represent the values of the reconstructed solution and its derivatives at the cell center, and $\Delta h$ denotes the average cell length. These unknown coefficients in Eq. (4.23) can be calculated by solving a least-squares problem with the mean values of surrounding cells [42].

Most of the time, the flow fields sampled from computational particles usually suffer from statistical noise, which may influence the accuracy and stability of the above interpolation algorithm. Alternatively, space interpolation based on the random motion of the computational particles can be employed. The details of this algorithm, which also yields second spatial order, are found in ref. [41]. To calculate the macro quantity $\Phi^{(k)}(\mathbf{x},t)$ of the kth particle at location $\mathbf{x}$ and time t, this interpolation algorithm is implemented as follows:

(a) A random deviation $\Delta\mathbf{x}$ is first sampled based on symmetric kernel function, e.g., a uniform hat kernel

$$K_{Uniform}(\Delta\mathbf{x}) = \begin{cases} 1 & \Delta x_i \in [-\Delta h/2, \Delta h/2] \\ 0 & otherwise \end{cases}, \tag{4.24}$$

where $\Delta x_i = \Delta h \cdot (rand - 0.5)$ is calculated, and $rand$ is a uniform random number in $[0,1]$.

(b) If the sampled random new position of particle k is located in cell j, i.e., if $\mathbf{x} + \Delta\mathbf{x} \in \Omega_j$, then the interpolated value of the particle k is set to

$$\Phi^{(k)}(\mathbf{x},t) \equiv \Phi_c(\mathbf{x}_j, t; \mathbf{x} + \Delta\mathbf{x} \in \Omega_j), \tag{4.25}$$

where $\Phi_c(\mathbf{x}_j, t)$ denotes the mean value in cell j.

In addition, since the USP-BGK method and DSMC track molecular motion in a similar way, the same treatment of the boundary conditions can be employed.

4.5 Conservation of the USP-BGK method

As well known, stochastic particle BGK methods can only ensure the conservation laws in a statistical sense. Therefore, energy and moment conservation correction steps are performed at the end of the relaxation step. In theory, according to Eqs. (C9)-(C11), translational ( $\hat{E}_{tr}$ ), rotational ( $\hat{E}_{rot}$ ) and vibrational ( $\hat{E}_{vib}$ ) energies based on $\hat{\mathcal{F}}(\mathbf{V}, I_r, I_v; \mathbf{x}, t^{n+1})$ are determined by the variables before the relaxation step. However, in practice, due to the nature of the random process, these energies are not exactly conserved after the relaxation step. To make them conservative, a correction step for the individual particle (similar to that in ref. [40]) is implemented as follows.

For the vibrational energy, a modification factor $\alpha_{vib} = \hat{E}_{vib}/\hat{E}'_{vib}$ is first calculated, where $\hat{E}'_{vib}$ denotes the vibrational energy directly averaged from the computational particles after the relaxation step. Then the vibrational energy level is corrected as

$$i^{(k)} = \text{int}\left(\alpha_{vib} i^{(k)'} + Rand\right), \tag{4.26}$$

where $i^{(k)}$ and $i^{(k)'}$ are the vibrational energy levels of the kth particle after and before modification, the operator $\text{int}$ denotes the integer part and $Rand$ is a uniform deviate in $[0,1]$.

Assuming the sum of the translational and vibrational energies are conserved, the modification factor for the translational energy can be calculated as

$$\alpha_{tr} = \sqrt{\left(\hat{E}_{tr} + \hat{E}_{vib} - \sum_{k=1}^{N} i^{(k)} RT_0 / N\right) / \hat{E}_{tr}'}. \tag{4.27}$$

Similarly, $\hat{E}_{tr}'$ denotes the translational energy averaged from the computational particles before correction. Then the particle velocity is finally modified as

$$\mathbf{M}^{(k)} = \alpha_{tr}\left(\mathbf{M}^{(k)'} - \sum_{k=1}^{N} \mathbf{M}^{(k)'} / N\right) + \mathbf{U}^*, \tag{4.28}$$

where $\mathbf{U}^*$ is the mean velocity before the relaxation step, which is constant during the relaxation step. $\mathbf{M}^{(k)}$ and $\mathbf{M}^{(k)'}$ are the velocities of the kth particle after and before correction, respectively. Conservation of momentum is also ensured by Eq. (4.28).

In addition, the particle's rotational energy is modified as

$$I_r^{(k)} = \alpha_{rot} I_r^{(k)'}, \text{ and } \alpha_{rot} = \hat{E}_{rot} / \hat{E}_{rot}'. \tag{4.29}$$

where $\hat{E}_{rot}'$ denotes the rotational energy directly averaged from the computational particles before correction. $I_r^{(k)}$ and $I_r^{(k)'}$ are the rotational energies of the kth particle after and before modification, respectively.

## 5. The hybrid USPBGK-DSMC method for polyatomic gases

5.1 The hybrid algorithm

Since the ES-BGK model is only acceptable at small Knudsen numbers, therefore, a hybrid scheme combining USP-BGK and DSMC performs better for multiscale gas flows. The hybrid USPBGK-DSMC method for monatomic gases was proposed recently [29]. For polyatomic gases, this hybrid scheme can be implemented in the same manner by replacing the BGK solver with the presented USP-BGK method for polyatomic gases. Based on Eq. (4.9) the governing equations of the hybrid method are

**transport step:** $\quad \frac{\partial \mathcal{F}}{\partial t} + V_i \frac{\partial \mathcal{F}}{\partial x_i} = P_c Q_C(\mathcal{F})$ and $\quad$ (5.1a)

**relaxation step:** $\quad \frac{\partial \mathcal{F}}{\partial t} = (1 - P_c) J_{(\text{Boltzmann})} + P_c Q_R(\mathcal{F})$, $\quad$ (5.1b)

where $P_c$ is a switching parameter between DSMC and USP-BGK. If $P_c = 1$, Eq. (5.1) reduces to Eq. (4.9), which corresponds to the USP-BGK method. Otherwise, if $P_c = 0$, Eq. (5.1) recovers the governing equations of the splitting scheme for the Boltzmann equation, e.g. DSMC. In practice, the choice of $P_c$ should be related to the Knudsen number. In the present work, we directly employ the criterion proposed in ref. [29], i.e.,

$$P_c = \begin{cases} 1 & \varepsilon / \Delta t \leq 1.5 \\ 0 & \varepsilon / \Delta t > 1.5 \end{cases}.$$
(5.2)

This criterion involves the Knudsen number and also considers the balance of accuracy and efficiency.

After determining the switching parameter $P_c$, similar to monatomic gas flow, one can implement the hybrid USPBGK-DSMC method for polyatomic gases. The detailed implementation is described in Table 3.

Table 3. Outline of the implementation of the hybrid USPBGK-DSMC method

| | |
|---|---|
| 1. **Initialization** | Introduce initial computational particles in the computational domain (similar as in DSMC). |
| 2. **Transport** | Move particles with their velocities and apply boundary conditions (similar as in DSMC). |
| 3. **Relaxation** | If $P_c = 1$, the particle velocities are updated according to the USP-BGK method (see Table 2); if $P_c = 0$, the DSMC collision operator is applied. |
| 4. **Sampling** | Sample the macroscopic quantities (similar as in DSMC). For the computational cells in which the USP-BGK method was employed, |

sampling is calculated from the auxiliary PDFs as shown in Appendix C.

5.2 Determining the rotational and vibrational collision numbers in the hybrid scheme

In the ES-BGK model [22] the rotational and vibrational energies relax to the intermediate translational-rotational temperature $T_{tr,rot}$ and equilibrium temperature $T_{eq}$, respectively. Using Eq. (A1), their relaxation processes are described as

$$\frac{\partial E_{vib}}{\partial t} = -\frac{1}{Z_{vib}\tau_{BGK}}\left(E_{vib} - e_{vib}(T_{eq})\right) \quad \text{and} \tag{5.3a}$$

$$\frac{\partial \left(E_{rot} - e_{rot}(T_{tr,rot})\right)}{\partial t} = -\frac{1}{Z_{rot}\tau_{BGK}}\left(E_{rot} - e_{rot}(T_{tr,rot})\right). \tag{5.3b}$$

In contrast, most continuum analyses use Jeans' equation to model rotational relaxation and the Landau-Teller equation to model vibrational relaxation [43]. To determine the rotational and vibrational collision numbers of the ES-BGK model, i.e., $Z_{rot}$ and $Z_{vib}$, it is necessary to find their relationship with the continuum model, i.e., with $Z_{rot}^{cont}$ and $Z_{vib}^{cont}$. For the continuum model, the relaxation of the rotational and vibrational energies can be written as

$$\frac{\partial E_{int}}{\partial t} = -\frac{1}{Z_{int}^{cont}\tau_c}\left(E_{int} - e_{int}(T_{tr})\right), \tag{5.4}$$

where the subscript "int" can be "rot" or "vib" refer to rotational and vibrational relaxation, respectively, and $\tau_c$ is the mean collision time. Comparing Eqs. (5.3a) and (5.4) yields

$$Z_{vib} = Z_{vib}^{cont}\frac{\tau_c}{\tau_{BGK}}\frac{e_{vib}(T_{vib}) - e_{vib}(T_{eq})}{e_{vib}(T_{vib}) - e_{vib}(T_{tr})}. \tag{5.5}$$

Similarly, comparing Eqs. (5.3b) and (5.4) yields

$$Z_{rot} = Z_{rot}^{cont}\frac{\tau_c}{\tau_{BGK}}\frac{e_{rot}(T_{rot}) - e_{rot}(T_{tr,rot}) + \frac{Z_{rot}}{Z_{vib}}\left(e_{rot}(T_{tr,rot}) - e_{rot}(T_{eq})\right)}{e_{rot}(T_{rot}) - e_{rot}(T_{tr})}. \tag{5.6}$$

Assuming that rotational relaxation is much faster than vibrational relaxation, Eq. (5.6)

can be further simplified as

$$Z_{rot} = Z_{rot}^{cont} \frac{\tau_c}{\tau_{BGK}} \frac{3}{3+\delta_{rot}}. \tag{5.7}$$

For the DSMC method the Borgnakke-Larsen model is employed in the present paper, and there is a similar relationship for rotational and vibrational collision numbers between DSMC and the continuum model. Here we directly use the results provided by refs. [44, 45], i.e.,

$$Z_{vib}^{DSMC} = Z_{vib}^{cont} \frac{\delta_T}{\delta_T + \delta_A} \quad \text{and} \quad Z_{rot}^{DSMC} = Z_{rot}^{cont} \frac{\delta_T}{\delta_T + \delta_{rot}}, \tag{5.8}$$

where $\delta_A = \delta_{vib}^2(T_{tr})\exp(T_0/T_{tr})/2$ (see ref. [44]) and $\delta_T = 4 - 2(\omega - 0.5)$ is the degrees of freedom of the relative translational energy. $\omega$ is the power-law exponent of the viscosity-temperature dependence, i.e., the viscosity relates to the temperature by the power-law

$$\mu = \mu_{ref}\left(T_{tr}/T_{ref}\right)^{\omega}, \tag{5.9}$$

where $T_{ref}$ and $\mu_{ref}$ are the reference temperature and viscosity, respectively.

In particular, for nitrogen gas, which is considered for the numerical cases presented in the next section, the rotational collision number of the continuum model is obtained as [46]

$$Z_{rot}^{cont} = Z_{rot}^{\infty} \bigg/ \left[1 + \frac{\pi^{3/2}}{2}\left(\frac{T^*}{T_{tr}}\right)^{1/2} + \left(\frac{\pi^2}{4} + \pi\right)\left(\frac{T^*}{T_{tr}}\right)\right], \tag{5.10}$$

where $Z_{rot}^{\infty} = 18.1$ and $T^* = 91.5K$. The vibrational collision number is set to

$$Z_{vib}^{cont} = \frac{\tau_{MW} + \tau_p}{\tau_c}, \tag{5.11}$$

where Millikan-White correlated vibrational relaxation time $\tau_{MW}$ is written as

$$\tau_{MW} = \frac{1}{p}\exp\left[\left(AT_{tr}^{-1/3} - B\right) - 18.42\right] \tag{5.12}$$

with $A = 221$ and $B = 12.3$ for nitrogen. With Park's high-temperature correction one obtains

$$\tau_p = \sqrt{\frac{\pi m}{8k_B T_{tr}}} \bigg/ (\sigma_s n), \tag{5.13}$$

where a reference collision cross-section of $\sigma_s = 5.81 \times 10^{-21} \, m^2$ is used.

## 6. Numerical simulations

In this section, four typical 1D and 2D gas flows were simulated and analyzed. For steady flows exponentially weighted moving time averaging [9] is employed to reduce statistical noise. Thus the macro variables $\Phi^{n+1}$ are calculated as

$$\Phi^{n+1} = \frac{n_a - 1}{n_a} \Phi^n + \frac{1}{n_a} \Phi(t), \tag{6.1}$$

where $n_a$ is the number of time steps used for averaging (in this paper we set $n_a = 10$) and $\Phi(t)$ denotes the macro quantities sampled at the current time.

6.1 Rotational and vibrational relaxation in a homogeneous flow

The proposed USP-BGK method for polyatomic gases first was tested for the relaxation problem in a homogeneous nitrogen flow. For DSMC the variable-hard-sphere (VHS) model was employed with a reference diameter of $4.17 \times 10^{-10} \, m$ at $T_{ref} = 273K$. The reference viscosity is $\mu_{ref} = 1.6734 \times 10^{-5} \, Pa \cdot s$ with a power-law exponent of $\omega = 0.75$, the initial translational temperature is 5000 K, and both initial rotational and vibrational temperatures are 0 K. The rotational and vibrational collision numbers are assumed to be constant, i.e., $Z_{rot}^{cont} = 5$ and $Z_{vib}^{cont} = 10$. According to the ES-BGK model, the analytical solutions of translational, rotational and vibrational temperature relaxations can be calculated from Eqs. (A2) and (A3). Figure 1 compares the relaxations obtained by DSMC, USP-BGK and from the analytical solution of the ES-BGK model. The time step sizes of DSMC and USP-BGK are $0.2\tau_c$ and $10\tau_c$, respectively. As shown in Fig.1 the USP-BGK method provides the correct relaxation rate of internal energies. Note that DSMC (dashed lines) and analytical solutions of the translational temperature (solid lines) lie on top of each other.

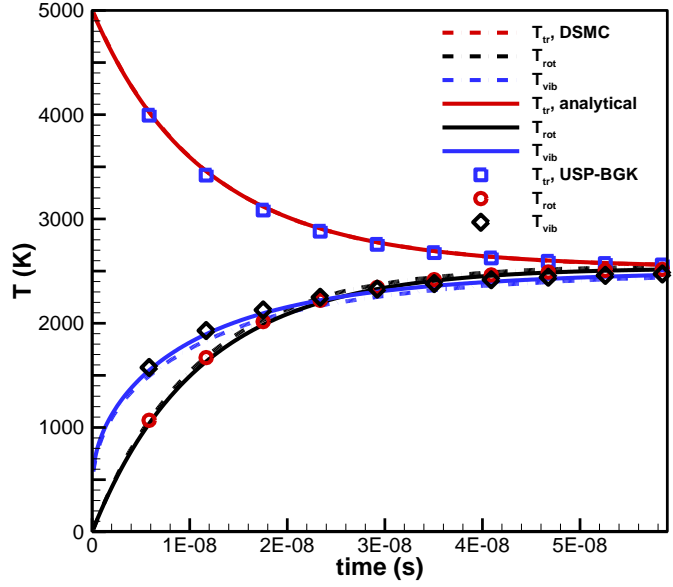

Figure 1. Comparison of rotational and vibrational relaxation in a homogeneous flow.

6.2 Shock structure

Shock waves are typical nonequilibrium flows. The structures of translational, rotational and vibrational temperatures usually do not overlap due to their distinguished relaxation rates. In the current case, where nitrogen is considered, the pre-shock temperature, number density and Mach number are $T_1 = 226.149K$, $n_1 = 3.745 \times 10^{23} m^{-3}$ and $Ma_1 = 15$, respectively; and in the post-shock side $T_2 = 8295.14K$, $n_2 = 2.793 \times 10^{24} m^{-3}$ and $Ma_2 = 0.344$. Note that the shock condition for two specific heat ratios can be found in ref. [47]. The VHS model was employed in DSMC with a reference diameter of $4.17 \times 10^{-10} m$, $T_{ref} = 273K$ and a viscosity exponent of 0.75. The rotational and vibrational relaxation numbers are assumed to be constant, i.e., $Z_{rot}^{cont} = 5$ and $Z_{vib}^{cont} = 20$. The size of the computational domain is $100\lambda$, where $\lambda$ is the mean free path length on the pre-shock side. For DSMC 100 uniform cells were employed, and for the hybrid USPBGK-DSMC method 50 uniform cells were used. The time step size was determined by the CFL criterion based on the pre-shock condition resulting in $CFL = 0.5$ for both methods.

Density, translational, rotational and vibrational temperature structures are shown in Figure 2. It can be observed that the results of DSMC and USPBGK-DSMC agree well. The blue line in Fig. 2(a) refers to the switching parameter of Eq. (5.2), which indicates the separation of the computational domain. Ahead of the shock wave, the gas flow is far from equilibrium, and thus the hybrid method switched to DSMC.

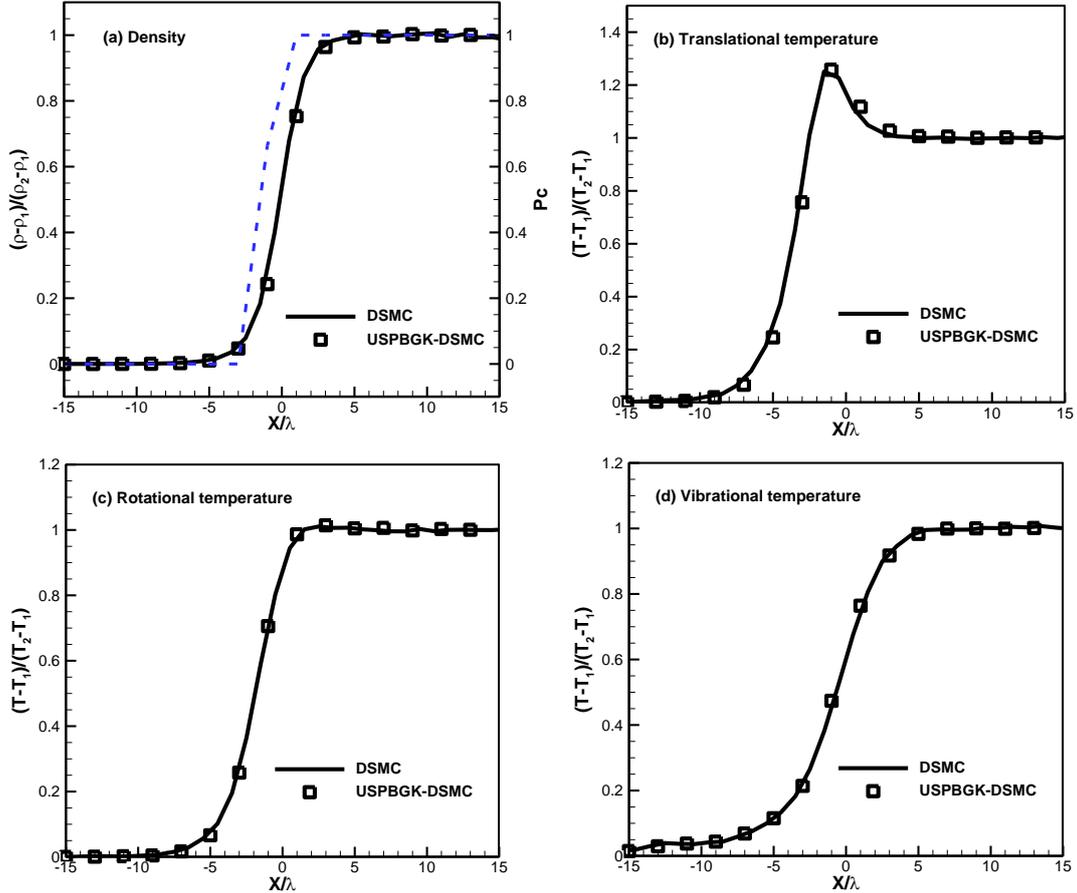

Figure 2. Comparison of density, translational, rotational and vibrational temperatures across the shock wave.

6.3 Hypersonic flow past a wedge

Hypersonic flow passing across a wedge with a sharp leading edge was calculated and is presented here. The flow is characterized by an attached shock and nonequilibrium regions near the leading edge, in the boundary layer and the wake. The computational setup is the same as in reference [48]. The wedge half-angle is 10-deg and the computational domain is a rectangular box of $0.8 \times 2.5\ m^2$; see the schematic in Fig. 3, which also depicts the unstructured mesh. About 27,000 cells and $CFL = 0.25$ were employed, and the average number of computational particles per cell was 200. Nitrogen inflow with a number density of $n_\infty = 2.124 \times 10^{21} \, m^{-3}$ at a temperature of $T_\infty = 200 K$ is considered, and the freestream velocity is $2883 \, m/s$ (Ma=10). The wedge wall temperature is 500 K and a fully diffusive wall condition is employed. The Knudsen number, calculated based on the freestream condition and the half-height of the wedge, is about 0.002. The bottom surface is treated as symmetric boundary, while the other surfaces of the computational domain are inflow and outflow boundaries, respectively. Rotational and vibrational collision numbers were set according to Eqs. (5.10)-(5.13). For DSMC the variable-hard-sphere (VHS) model was used with a reference diameter of $4.11 \times 10^{-10} \, m$, the reference temperature is $T_{ref} = 290K$ with a power-law exponent of $\omega = 0.7$.

Figure 4(a) shows the distributions of Mach number and the separation of USP-BGK and DSMC regions employed for the calculation. It can be observed that the gas is very rarefied in the wake region; thus DSMC was employed there, while the USP-BGK method was used in the remaining domain. Figure 4(b) presents the translational and rotational temperatures. Since the Knudsen number in the current case is small, translational and rotational relaxations are in equilibrium in most parts. However, minor differences can still be observed in the wake of the wedge due to the rarefication effect. In addition, the vibrational energy in the present case is hardly excited due to low-temperature values.

Surface friction ($C_F$) and heating ($C_H$) coefficients are compared in Fig. 5. These dimensionless coefficients are defined as

$$C_F = \sigma_{sur} / \left(\frac{1}{2}\rho_\infty U_\infty^2\right) \text{ and} \tag{6.2}$$

$$C_H = q_{sur} / \left(\frac{1}{2}\rho_\infty U_\infty^3\right), \tag{6.3}$$

where $\sigma_{sur}$ and $q_{sur}$ denote shear stress and heat flux on the surface, respectively. Further, $\rho_\infty$ is the free-stream density and $U_\infty$ is the free-stream velocity. The surface coefficients are plotted as functions of the distance S along the wedge surface normalized by the length L of the top surface. We observe that the USPBGK-DSMC results are virtually identical to the DSMC reference data of [49].

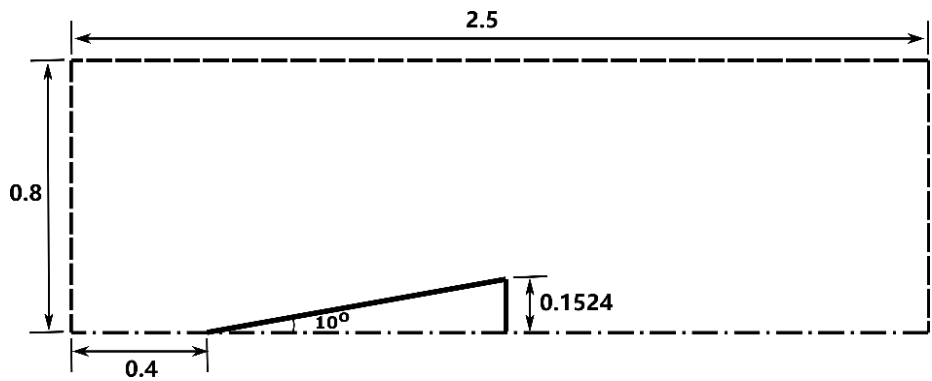

(a)

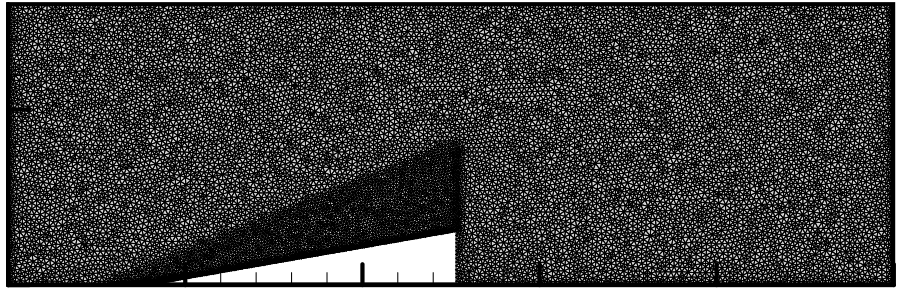

(b)

Figure 3. (a) Illustration of the wedge test case with (b) the computational grid.

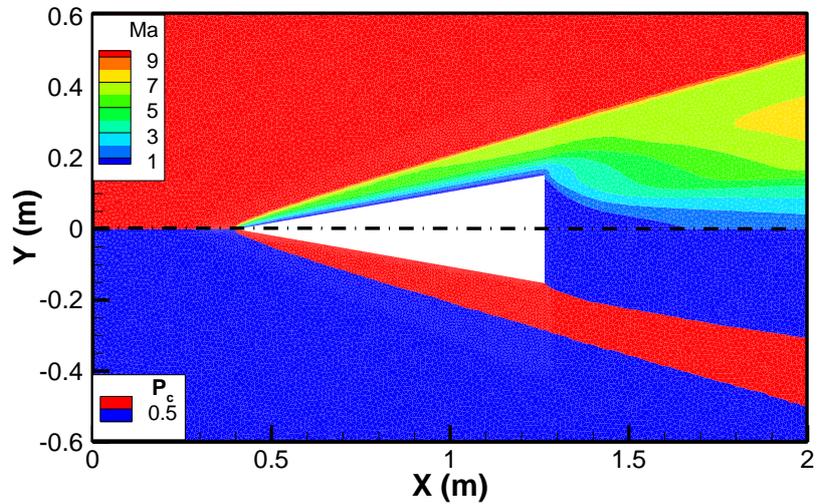

(a)

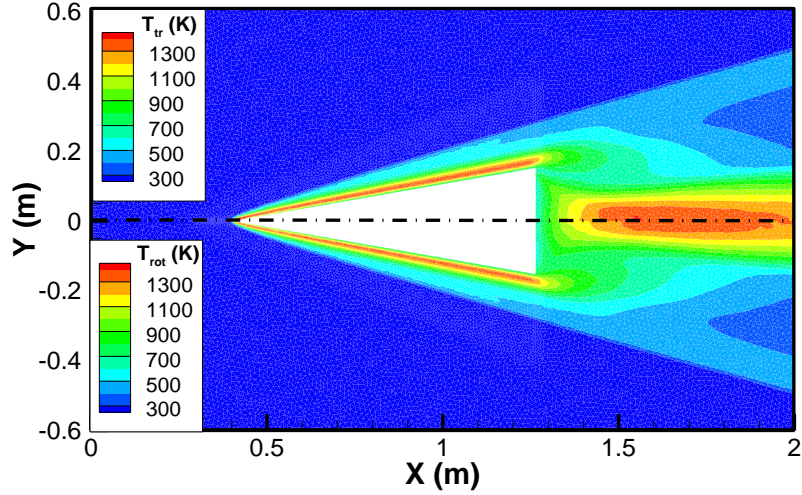

(b)

Figure 4. Distribution of (a) Mach number and switching parameter; (b) the translational and rotational temperatures for a Mach 10 nitrogen flow across a wedge.

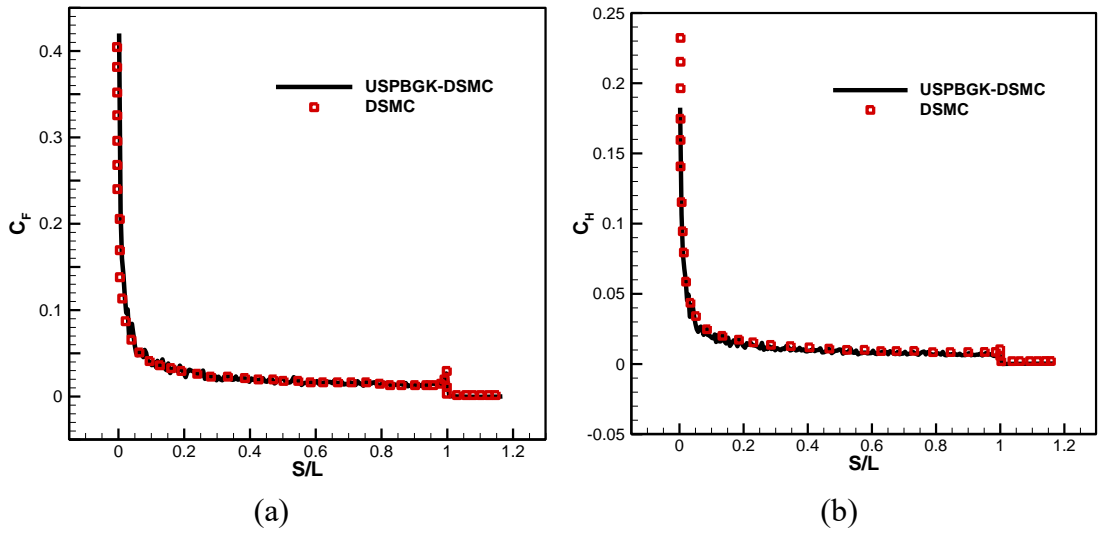

Figure 5. (a) Surface friction and (b) heating coefficients for a Mach 10 nitrogen flow across a wedge. The solid lines represent the USPBGK-DSMC results and the square symbols represent the DSMC data given in ref. [49].

6.4 Hypersonic flow past a cylinder

Hypersonic flow past a cylinder is another typical case to investigate multiscale gas flows. A high-density continuum region is found in front of the body and a rarefied region in its wake. In this paper, the same computational parameters as used in [46] are employed to validate the proposed hybrid method. Nitrogen flow with a freestream number density of $n_\infty = 1.61 \times 10^{21}\ m^{-3}$ and initial temperatures of $T_{tr,\infty} = T_{rot,\infty} = T_{vib,\infty} = 217.5 K$ is considered. The freestream velocity is $4510 \, m/s$ (Ma=15) and the cylinder diameter is 0.08 m. A fully diffusive wall condition is employed, while the surface temperature of the cylinder is kept at $1000 K$. For DSMC the variable-hard-sphere (VHS) model is employed with a reference diameter of $4.17 \times 10^{-10} m$ at $T_{ref} = 273K$ with a power-law exponent of $\omega = 0.75$. The rotational and vibrational collision numbers are set according to Eqs. (5.10)-(5.13). The Knudsen number, calculated based on the free-stream condition and the cylinder diameter, is equal to 0.01.

The computational domain is a rectangular box of $0.3 \times 0.15 \, m^2$ as shown in Fig. 6. The left and bottom sides are inlet and symmetry boundaries, respectively, while the other sides are outlet boundaries. Cartesian grids with successive refinement near the cylinder surface were employed. For the coarsest level, resolutions of $36 \times 18$, $72 \times 36$ and $144 \times 72$ were used. The time step size was determined by the CFL condition with $CFL = 0.5$. For all simulations, the average number of computational particles per coarse cell was 200.

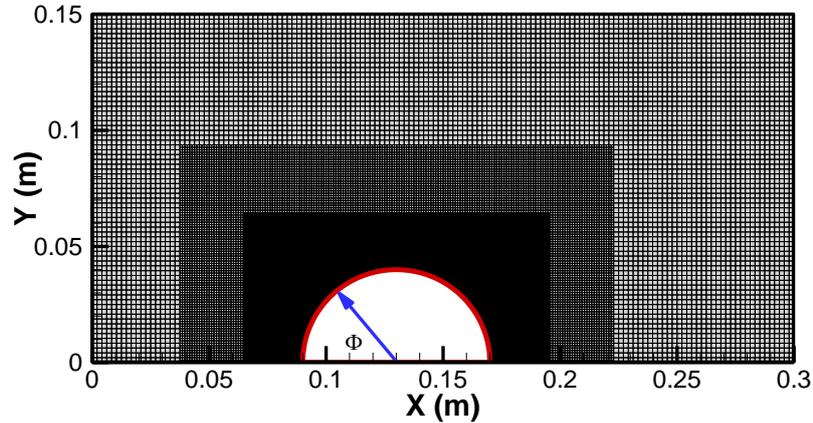

Figure 6. Computational domain with successively refined grid for hypersonic flow past a cylinder.

Figure 7 shows the distributions of Mach number, translational and vibrational temperatures. The lower picture of Fig. 7(a) depicts the switching parameter $P_c$; one can observe continuum regime in front of the cylinder where $P_c = 1$ and thus the USP-BGK method was employed; in the remaining rarefied region, DSMC was used. Figure 8 compares translational, rotational and vibrational temperatures along an extraction line at an angle of 45°. The results of the USPBGK-DSMC method agree well with the DSMC data given in ref. [46]. The black dash-dotted line indicates the interface between USP-BGK and DSMC regions. It can be observed that the post-shock area lies in the continuum regime.

The surface heating coefficient is compared in Figure 9. For the finest grid, the USPBGK-DSMC results are virtually identical to the DSMC reference data of [46]. However, the traditional BGK-DSMC method, which is less accurate in the continuum regime, converges slower to the exact solution as the number of computational cells increases. Note that instead of the USP-BGK method, the stochastic particle BGK solver introduced in section 3 was employed in the traditional BGK-DSMC method.

The L1-norm errors of the surface heating coefficient for the different grid sizes are given in Table 4. Note that the order of convergence for the USPBGK-DSMC method is about 2.5, which is larger than that for the traditional BGK-DSMC method. All simulations were carried out on a single Intel Xeon E5-2680v3 CPU of ETH's "Euler II" cluster. The CPU time per time step also is presented. It can be seen that the computational times are similar if the same cell sizes are used. However, due to the higher order of accuracy, the USPBGK-DSMC method requires fewer cells and can employ larger time steps to achieve the same level of accuracy. Therefore the computational cost is reduced significantly.

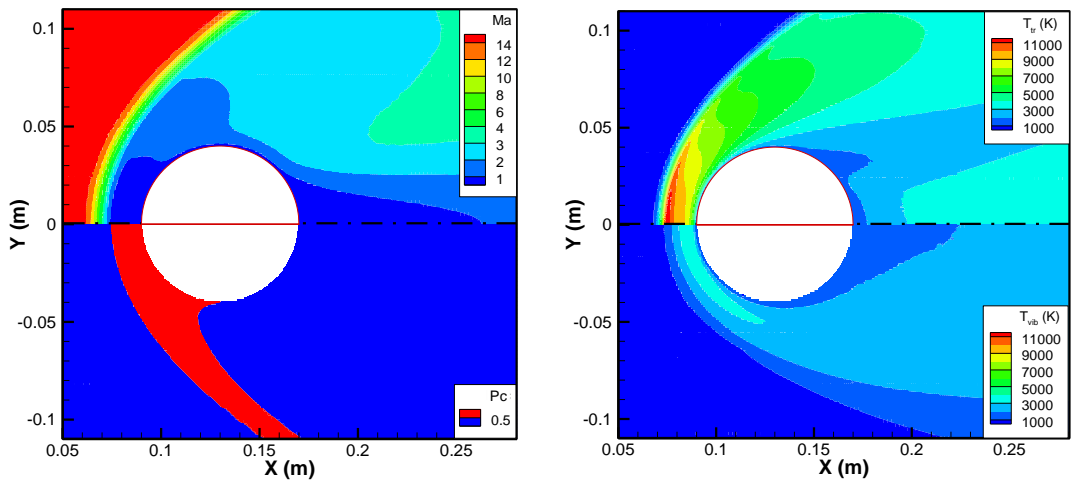

(a)                                (b)

Figure 7. Nitrogen flow around a cylinder at a Mach number of 15: Distribution of (a) Mach number and switching parameter; (b) translational and vibrational temperatures. To obtain these results the USPBGK-DSMC method was employed with a coarse grid of $144 \times 72$.

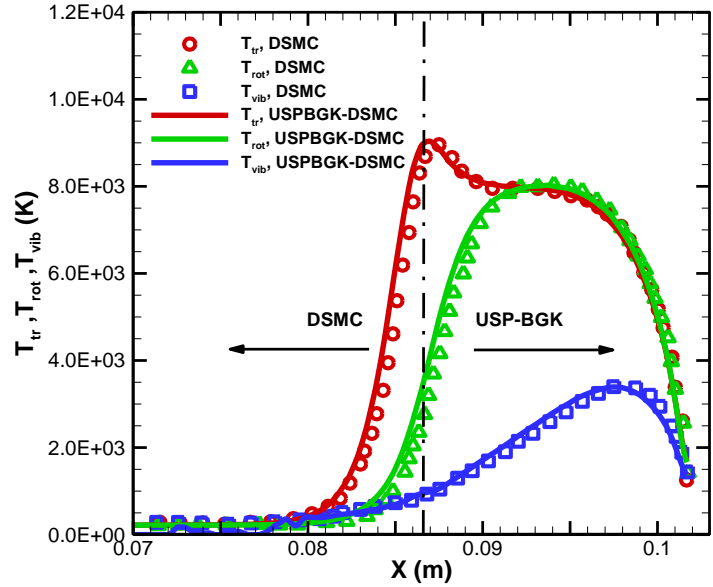

Figure 8. Nitrogen flow around a cylinder at a Mach number of 15 and Kn=0.01: Temperature distributions along the 45° extraction line. The symbols indicate DSMC [46] and the solid line USPBGK-DSMC results. The black dash-dotted line refers to the interface between USPBGK and DSMC.

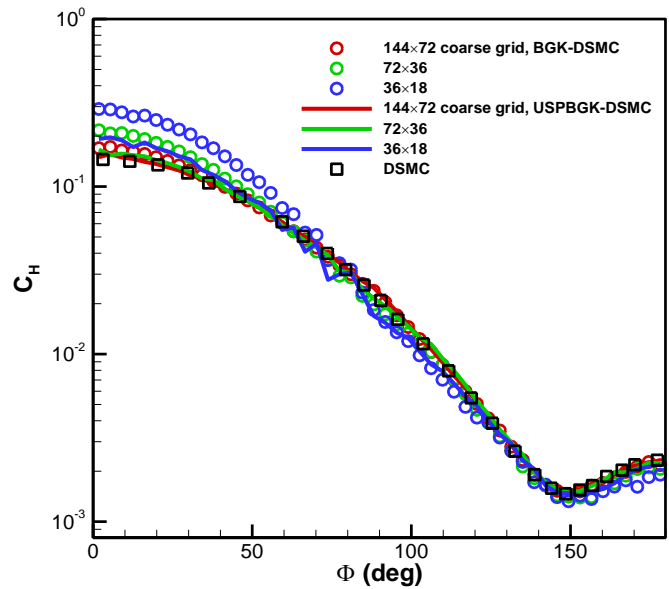

Figure 9. Nitrogen flow around a cylinder at a Mach number of 15: Surface heating

coefficient along the cylinder wall. The solid lines represent USPBGK-DSMC results, the circle symbols traditional BGK-DSMC results, and the square symbols DSMC data of [46]. For both hybrid methods, three simulations with grids of different resolutions were performed, i.e., respective base coarse level grids of 144 × 72, 72 × 36 and 36 × 18 cells were employed.

**Table 4**. $L_1$-norm error of surface heating coefficient along the cylinder wall.

| Number of coarse cells | BGK-DSMC | | | USPBGK-DSMC | | |
|---|---|---|---|---|---|---|
| | $L_1$-norm error | Order | CPU time (s) | $L_1$-norm error | Order | CPU time(s) |
| 36×18 | 0.5730 | — | 0.010 | 0.1288 | — | 0.017 |
| 72×36 | 0.2053 | 1.48 | 0.026 | 0.0214 | 2.59 | 0.035 |

## 7. Conclusion

Employing the micro-macro decomposition of the ES-BGK collision term given in ref. [22], the USP-BGK method was extended for applications of polyatomic gas flows. Since the simplified algorithm of the USP-BGK method can be implemented in the same way as the traditional SP-BGK method, we have shown that constructing a hybrid method combining the USP-BGK and DSMC methods is straightforward. A proper switching criterion and consistent rotational and vibrational collision numbers were determined and have been validated by the simulations of supersonic flows with strong shocks. Hypersonic flows past a wedge and a cylinder were calculated by the hybrid USPBGK-DSMC method, and it has been demonstrated that the results agree well with the corresponding DSMC solutions. Note that for both test cases, a huge Knudsen number range from continuum to high rarification has to be modelled. Compared to the traditional stochastic particle methods, the efficiency of the presented hybrid method is excellent and the computational cost is much lower. Therefore the hybrid USPBGK-DSMC method is suitable for simulations of complex multiscale flows for polyatomic gases.

## Acknowledgements

This work was supported by the National Natural Science Foundation of China (No. 51876076) and the LHD Youth Innovation Fund from the State Key Laboratory of High Temperature Gas Dynamics (Grant No. LHD2019CX12).

## Appendix A Relaxation of the ES-BGK model

Note that the density $\rho(t)$, the mean velocity $\mathbf{U}(t)$ and the equilibrium temperature $T_{eq}(t)$ are constant during the relaxation step, while $\mathbf{\Pi}(t)$, $E_{tr}(t)$, $E_{rot}(t)$ and $E_{vib}(t)$ can be obtained by solving the moment equation of Eq. (3.1b).

Multiplying Eq. (3.1b) by $(V-U)^2/2$, $I_r$ and $iRT_0$, and integrating over the phase space, we obtain

$$\frac{\partial}{\partial t}\begin{pmatrix} E_{tr}(t) \\ E_{rot}(t) \\ E_{vib}(t) \end{pmatrix} = \frac{\eta}{\tau_{BGK}}\begin{pmatrix} e_{tr}(T_{eq}) - E_{tr}(t) \\ e_{rot}(T_{eq}) - E_{rot}(t) \\ e_{vib}(T_{eq}) - E_{vib}(t) \end{pmatrix} + \frac{(1-\eta)}{\tau_{BGK}}\begin{pmatrix} -\dfrac{\delta_{rot}\theta}{3+\delta_{rot}} & \dfrac{3\theta}{3+\delta_{rot}} & 0 \\ \dfrac{\delta_{rot}\theta}{3+\delta_{rot}} & -\dfrac{3\theta}{3+\delta_{rot}} & 0 \\ 0 & 0 & 0 \end{pmatrix}\begin{pmatrix} E_{tr}(t) \\ E_{rot}(t) \\ E_{vib}(t) \end{pmatrix}. \quad (A1)$$

Assuming $a_1 = E_{tr}+E_{rot}$ and $a_2 = (\delta_{rot}/3)E_{tr} - E_{rot}$, the solution of Eq. (A1) is obtained as

$$a_1(t) = \exp\left(-\frac{\eta t}{\tau_{BGK}}\right)a_1(0) + \left[1-\exp\left(-\frac{\eta t}{\tau_{BGK}}\right)\right]\left(e_{tr}(T_{eq}) + e_{rot}(T_{eq})\right), \quad (A2.a)$$

$$a_2(t) = \exp\left[-\frac{((1-\eta)\theta+\eta)t}{\tau_{BGK}}\right]a_2(0), \quad (A2.b)$$

$$E_{vib}(t) = \exp\left(-\frac{\eta t}{\tau_{BGK}}\right)E_{vib}(0) + \left[1-\exp\left(-\frac{\eta t}{\tau_{BGK}}\right)\right]e_{vib}(T_{eq}). \quad (A2.c)$$

Therefore,

$$E_{tr}(t) = \frac{3}{3+\delta_{rot}}(a_1+a_2) \quad \text{and} \quad E_{rot}(t) = \frac{\delta_{rot}a_1 - 3a_2}{3+\delta_{rot}}. \quad (A3)$$

Besides $E_{tr}(t)$ and $E_{rot}(t)$, $\mathbf{\Pi}(t)$ is depended on the shear stress $\mathbf{\sigma} = \Theta - RT_{tr}\mathbf{I}$.

Multiplying Eq. (3.1b) by $(\mathbf{V}-\mathbf{U})\otimes(\mathbf{V}-\mathbf{U})-(V-U)^2/3$ and integrating over the phase space, we have

$$\frac{\partial \boldsymbol{\sigma}(t)}{\partial t} = -\frac{1}{\tau_{BGK}}\left[1-(1-\eta)(1-\theta)\nu\right]\boldsymbol{\sigma}(t). \tag{A4}$$

Therefore,

$$\boldsymbol{\sigma}(t) = e^{-t/(\tau_{BGK}\mathrm{Pr})}\boldsymbol{\sigma}(0). \tag{A5}$$

According to Eq. (2.9), $\boldsymbol{\Pi}(t)$ reads

$$\boldsymbol{\Pi}(t) = \eta RT_{eq}\mathbf{I} + (1-\eta)\left[\theta RT_{tr,rot}(t)\mathbf{I} + (1-\theta)\left(\nu\boldsymbol{\sigma}(t) + RT_{tr}(t)\mathbf{I}\right)\right], \tag{A6}$$

## Appendix B Metropolis-Hastings (MH) algorithm

A Metropolis-Hastings method produces a Markov chain of samples from the target distribution [50] and has been employed to sample Chapman-Enskog [51] and ES-BGK distribution [52]. We also use the Metropolis-Hastings algorithm to generate the target distribution in the present work. The basic idea of the Metropolis-Hastings algorithm is to design a Markov chain with a transition probability $p(\mathbf{X}_{old},\mathbf{X}_{new})$ between states, e.g. from $\mathbf{X}_{old}$ to $\mathbf{X}_{new}$. According to the reversibility condition, the transition probability should satisfy

$$f_t(\mathbf{X}_{old})p(\mathbf{X}_{old},\mathbf{X}_{new}) = f_t(\mathbf{X}_{new})p(\mathbf{X}_{new},\mathbf{X}_{old}), \tag{B1}$$

where $f_t(\mathbf{X})$ is the target distribution and $\mathbf{X}$ is a stochastic variable. A simple way to satisfy this condition is to take [51]

$$p(\mathbf{X}_{old},\mathbf{X}_{new}) = M(\mathbf{X}_{new})\begin{cases} 1 & \text{if } f_t(\mathbf{X}_{new})M(\mathbf{X}_{old}) \geq f_t(\mathbf{X}_{old})M(\mathbf{X}_{new}) \\ \dfrac{f_t(\mathbf{X}_{new})M(\mathbf{X}_{old})}{f_t(\mathbf{X}_{old})M(\mathbf{X}_{new})} & \text{if } f_t(\mathbf{X}_{new})M(\mathbf{X}_{old}) < f_t(\mathbf{X}_{old})M(\mathbf{X}_{new}) \end{cases}, \tag{B2}$$

where $M(\mathbf{X})$ can be chosen as a Maxwellian distribution, which is easy sampled. For example, to generate $f_t(\mathbf{X}) = \mathcal{F}_G$, we assume $M(\mathbf{X}) = \mathcal{F}_M$ given by Eq. (4.5), then the Metropolis-Hastings algorithm was implemented as follows:

(1) Draw $\mathbf{X}_{old} = \{\mathbf{V}_{old}, I_{r,old}, I_{v,old}\}$ from the Maxwellian distribution $\mathcal{F}_M$; select a new value if $\mathcal{F}_G(\mathbf{X}_{old}) < 0$.

(2) Draw a new value $\mathbf{X}_{new}$ from the Maxwellian distribution.

(3) Draw a random number $Rand$ between 0 and 1; if $Rand \leq \min\left(1, \dfrac{\mathcal{F}_G(\mathbf{X}_{new})\mathcal{F}_M(\mathbf{X}_{old})}{\mathcal{F}_G(\mathbf{X}_{old})\mathcal{F}_M(\mathbf{X}_{new})}\right)$ then accept the "move" and set $\mathbf{X}_{old} = \mathbf{X}_{new}$; otherwise keep the current value of $\mathbf{X}_{old}$.

(4) Repeat steps 2 and 3 until $N_{try}$ attempted moves have been made. Here, we set $N_{try} = 35$ as ref. [52].

Similarly, this implementation can also be employed to generate other distributions in this paper.

## Appendix C Macro variables averaged from the auxiliary distribution

First considering the auxiliary distribution $\tilde{\mathcal{F}}^* = \mathcal{F}^* - \dfrac{\Delta t}{2} Q_C^*$, since the continuum part of the collision term also satisfies the conservation law, therefore, for the conserved variables, such as the density, mean velocity and total energy, we have

$$\rho = \left\langle \tilde{\mathcal{F}}^* \right\rangle_{V, I_r, I_v} = \left\langle \mathcal{F}^* \right\rangle_{V, I_r, I_v}, \quad \rho\mathbf{U} = \left\langle \mathbf{V}\tilde{\mathcal{F}}^* \right\rangle_{V, I_r, I_v} = \left\langle \mathbf{V}\mathcal{F}^* \right\rangle_{V, I_r, I_v} \quad \text{and}$$

$$\rho E = \left\langle \left(\frac{1}{2}(V-U)^2 + I_r + I_v\right)\tilde{\mathcal{F}}^* \right\rangle_{V, I_r, I_v} = \left\langle \left(\frac{1}{2}(V-U)^2 + I_r + I_v\right)\mathcal{F}^* \right\rangle_{V, I_r, I_v}. \tag{C1}$$

For the shear stress, multiplying Eq. (4.12a) by $(\mathbf{V}-\mathbf{U})\otimes(\mathbf{V}-\mathbf{U}) - (V-U)^2/3$ and integrating over the phase space, we obtain

$$\boldsymbol{\sigma}^* = \tilde{\boldsymbol{\sigma}}^* \bigg/ \left(1 + \frac{\Delta t}{2\varepsilon \Pr}\right). \tag{C2}$$

For the translational, rotational and vibrational energies, multiplying Eq. (4.12a) by $\phi = \left(\dfrac{1}{2}(V-U)^2, I_r, I_v\right)^T$ and integrating over the phase space, we obtain

$$\tilde{E}_{tr}^* = E_{tr}^* - \frac{\Delta t}{2\varepsilon}\left[\eta\left(\frac{3}{2}RT_{eq} - E_{tr}^*\right) + (1-\eta)\left(\frac{3\theta}{3+\delta_{rot}} E_{rot}^* - \frac{\theta \delta_{rot}}{3+\delta_{rot}} E_{tr}^*\right)\right], \tag{C3}$$

$$\tilde{E}_{rot}^* = E_{rot}^* - \frac{\Delta t}{2\varepsilon}\left[\eta\left(\frac{\delta_{rot}}{2}RT_{eq} - E_{rot}^*\right) + (1-\eta)\left(-\frac{3\theta}{3+\delta_{rot}}E_{rot}^* + \frac{\theta\delta_{rot}}{3+\delta_{rot}}E_{tr}^*\right)\right], \quad (C4)$$

$$\tilde{E}_{vib}^* = E_{vib}^* - \frac{\Delta t}{2\varepsilon}\left[\eta\left(e_{vib}(T_{eq}) - E_{vib}^*\right)\right]. \quad (C5)$$

After organizing, the translational, rotational and vibrational energies after the transport step read

$$E_{tr}^* = \frac{\eta\frac{3\Delta t}{2\varepsilon}\frac{3+\delta_{rot}}{2}RT_{eq} + 3\left(\tilde{E}_{tr}^* + \tilde{E}_{rot}^*\right)}{\left(1+\frac{\Delta t}{2\varepsilon}\eta\right)(3+\delta_{rot})} + \frac{\left(\delta_{rot}\tilde{E}_{tr}^* - 3\tilde{E}_{rot}^*\right)}{\left(1+\frac{\Delta t}{2\varepsilon}(\eta+(1-\eta)\theta)\right)(3+\delta_{rot})}, \quad (C6)$$

$$E_{rot}^* = \frac{\eta\frac{\delta_{rot}\Delta t}{2\varepsilon}\frac{3+\delta_{rot}}{2}RT_{eq} + \delta_{rot}\left(\tilde{E}_{tr}^* + \tilde{E}_{rot}^*\right)}{\left(1+\frac{\Delta t}{2\varepsilon}\eta\right)(3+\delta_{rot})} - \frac{\left(\delta_{rot}\tilde{E}_{tr}^* - 3\tilde{E}_{rot}^*\right)}{\left(1+\frac{\Delta t}{2\varepsilon}(\eta+(1-\eta)\theta)\right)(3+\delta_{rot})}, \quad (C7)$$

$$E_{vib}^* = \frac{\tilde{E}_{vib}^* + \eta\frac{\Delta t}{2\varepsilon}e_{vib}(T_{eq})}{\left(1+\eta\frac{\Delta t}{2\varepsilon}\right)}. \quad (C8)$$

Similar results can be calculated from Eq. (4.12b), and the translational, rotational and vibrational energies after the relaxation step are averaged from the auxiliary distribution $\hat{\mathcal{F}}$. In addition, since $\langle\phi Q_C(\mathcal{F})\rangle_{V,I_r,I_v} = \langle\phi Q(\mathcal{F})\rangle_{V,I_r,I_v}$ as shown in Eq. (4.7), $E_{tr}^*$, $E_{rot}^*$ and $E_{vib}^*$ keep constant during the relaxation step. Therefore, $\hat{E}_{tr}$, $\hat{E}_{vib}$ and $\hat{E}_{vib}$ can also be calculated from the values before the relaxation step, i.e.,

$$\hat{E}_{tr} = E_{tr}^* + \frac{\Delta t}{2\varepsilon}\left[\eta\left(\frac{3}{2}RT_{eq} - E_{tr}^*\right) + (1-\eta)\left(\frac{3\theta}{3+\delta_{rot}}E_{rot}^* - \frac{\theta\delta_{rot}}{3+\delta_{rot}}E_{tr}^*\right)\right], \quad (C9)$$

$$\hat{E}_{vib} = E_{rot}^* + \frac{\Delta t}{2\varepsilon}\left[\eta\left(\frac{\delta_{rot}}{2}RT_{eq} - E_{rot}^*\right) + (1-\eta)\left(-\frac{3\theta}{3+\delta_{rot}}E_{rot}^* + \frac{\theta\delta_{rot}}{3+\delta_{rot}}E_{tr}^*\right)\right], \quad (C10)$$

$$\hat{E}_{vib} = E_{vib}^* + \frac{\Delta t}{2\varepsilon}\left[\eta\left(e_{vib}(T_{eq}) - E_{vib}^*\right)\right]. \quad (C11)$$

In the above, the hat "~" and "⌢" denote the variables averaged from $\tilde{\mathcal{F}}$ and $\hat{\mathcal{F}}$, respectively.

# Reference

[1] M.S. Ivanov, S.F. Gimelshein, Computational hypersonic rarefied flows, Annu.